\documentclass[aps,prd,twocolumn,superscriptaddress,floatfix,nofootinbib]{revtex4-2}

\usepackage{graphicx}
\usepackage{amsmath}
\usepackage{amssymb}
\usepackage{hyperref}
\usepackage{bm}
\usepackage{xcolor}

\graphicspath{{figures/}}

\newcommand{\HzToday}{66^{+44}_{-37}}
\newcommand{\HzOpt}{66^{+15}_{-13}}
\newcommand{\betaNow}{0.72^{+0.14}_{-0.16}}
\newcommand{\betaFut}{0.72\pm0.04}

\definecolor{pzcolor}{rgb}{0.94,0.05,0.05}

\definecolor{claudecolor}{rgb}{0.37,0.00,0.85}

\begin{document}
\rightline{FERMILAB-PUB-26-0752-T}

\raggedbottom  

\title{The road to $H_0$, $\Omega_m$, and $w$ with a global neutrino-telescope network}

\author{P.~Zhelnin}
\affiliation{Department of Physics and Laboratory for Particle Physics and
Cosmology, Harvard University, Cambridge, MA 02138, USA}

\author{G.~Herrera}
\affiliation{Department of Physics and Laboratory for Particle Physics and
Cosmology, Harvard University, Cambridge, MA 02138, USA}
\affiliation{Department of Physics and Kavli Institute for Astrophysics and
Space Research, Massachusetts Institute of Technology, Cambridge, MA 02139, USA}

\author{N.~Kamp}
\affiliation{Department of Physics and Laboratory for Particle Physics and
Cosmology, Harvard University, Cambridge, MA 02138, USA}

\author{C.~A.~Arg\"{u}elles}
\affiliation{Department of Physics and Laboratory for Particle Physics and
Cosmology, Harvard University, Cambridge, MA 02138, USA}

\date{\today}

\begin{abstract}

The newly measured correlation between the neutrino and X-ray luminosities of Seyfert galaxies has yielded the first neutrino-based measurement of the Hubble constant. 
Building on this result, we forecast the reach of a global neutrino-telescope network by 2030, utilizing full-sky IceCube exposures combined with KM3NeT-ARCA and Baikal-GVD.
We find that this 2030 configuration sharpens the neutrino-measured Hubble constant from today's $H_0 = \HzToday$ to $H_0 = \HzOpt$~km~s$^{-1}$~Mpc$^{-1}$. 
Additionally, we project that the disk--corona slope of the neutrino--X-ray luminosity relation ($L_\nu\propto L_X^\beta$) will reach $\beta=\betaFut$, enabling a $\geq \!5\sigma$ exclusion of the calorimetric value $\beta=1$. 
Finally, we show that, if a high-redshift population of neutrino sources, such as JWST's ``little red dots," is identified and obeys the same luminosity relation, we can obtain a new independent measurement of the matter density $\Omega_m$ and the dark energy equation-of-state parameter $w$.
\end{abstract}

\maketitle

\section{Introduction}
\label{sec:intro}

One of the sharpest disagreements in modern cosmology is the $\sim\!5\sigma$ tension~\cite{Verde:2019ivm,DiValentino:2021izs,Abdalla:2022yfr} between the Hubble constant inferred from the early universe by Planck, $67.4 \pm 0.5$~km~s$^{-1}$~Mpc$^{-1}$~\cite{Planck2018}, and the value measured locally by SH0ES, $73.0 \pm 1.0$~km~s$^{-1}$~Mpc$^{-1}$~\cite{Riess2022SH0ES}.
Resolving the origin of this tension, which could reflect an unrecognized systematic effect or new physics, calls for independent measurements of the expansion rate, each based on its own assumptions.

High-energy neutrinos from Seyfert galaxies offer one such measurement.
Recent work demonstrated that these sources exhibit a power-law relationship between their neutrino and X-ray luminosities, $L_\nu \propto L_X^\beta$~\cite{Herrera:2026now}.
This relationship, which was anticipated by disk--corona models~\cite{Murase:2019vdl,Inoue:2019yfs,Kheirandish:2021wkm,Murase:2022dog,Eichmann:2022lxh,Fiorillo:2023dts,Padovani:2024ibi,Murase:2022feu}, makes these sources standardizable candles, since the ratio between the neutrino luminosity, predicted by X-ray flux measurements, and the observed neutrino flux yields a distance.
Like any distance ladder, this method relies on ``anchors'', sources whose distances are known independently of $H_0$.
Because neutrinos travel unimpeded from their source to Earth, as do X-rays over Mpc distances, no in-transit systematics enter into this analysis method.
The method's systematics are therefore primarily confined to the source via the X-ray calibration of each Seyfert (Sec.~\ref{sec:xray}) and the
distances of the anchors. 
In their work, Herrera et al.~\cite{Herrera:2026now} extracted a first $H_0$ measurement from IceCube data which is consistent with both Planck and SH0ES and uncertain at the $\sim\!70\%$ level. 
Their work used only twelve northern-sky Seyferts, a single anchor, and one neutrino observatory.

In this work, we extend that initial study in three new directions.
First, we report a new measurement of $H_0$ using existing IceCube data, extending Ref.~\cite{Herrera:2026now} to include southern-sky Seyferts~\cite{Abbasi2026ESTES} and an additional anchor.
The measurement of $H_0$ achieved in this work, at $H_0 = \HzToday$, is consistent with the prior result.
Second, we forecast the sensitivity to $H_0$ by 2030, when a multi-detector network including KM3NeT-ARCA and Baikal-GVD will be operating~\cite{Spiering2026}, additional anchors will have been detected, and improved X-ray measurements of the anchor galaxies will be available.
Our central result in this configuration is that $H_0$ will sharpen from today's $\HzToday$ to $\HzOpt$~km~s$^{-1}$~Mpc$^{-1}$ and that the disk--corona slope $\beta$ will simultaneously be pinned to $\pm0.04$ (Sec.~\ref{sec:h0}). 
In this work, we do not consider the addition of any neutrino sources beyond those currently identified; the discovery of new Seyferts, especially in the mostly unexplored Southern sky, could further improve a 2030 $H_0$ measurement.

Finally, we also propose a new method for measuring $\Omega_m$ and $w$ with high energy neutrinos.
Specifically, we show that if the ``little red dots''~\cite{Akins2025LRD} observed by the James Webb Space Telescope (JWST), a potential high-redshift neutrino population \cite{KuzeMurase2026}, obey the same luminosity relation as Seyferts, extending the ladder to them would open a new window on the cosmological matter density, $\Omega_m$, and the dark-energy equation-of-state parameter, $w$ (Sec.~\ref{sec:de}). 

\section{The neutrino distance ladder}
\label{sec:method}

The expected number of signal neutrinos from a source $i$ is
\begin{equation}
  n_{s,i} = \kappa \, C_i \, F_{X,i}^{\,\beta}\, d_{L,i}^{\,2(\beta-1)}\,,
  \label{eq:model}
\end{equation}
where $F_{X,i}$ is the intrinsic $20$--$50$~keV X-ray flux~\cite{Ricci2017BASS}, $C_i$ the detector spectral response, $d_{L,i}$ the luminosity distance, and $\kappa$ an overall normalization (converting fluxes to luminosities recovers $L_\nu = \kappa' L_X^\beta$). 
A derivation of this relation is given in Ref.~\cite{Herrera:2026now}. 
Because the slope parameter $\beta$ is sub-linear ($\beta\simeq0.72\neq1$), the predicted signal count depends on distance, and hence on $H_0$, through the factor $d_L^{2(\beta-1)}$. 
For low-redshift Seyferts $d_L\simeq cz/H_0$, so $H_0$ enters as an overall scale degenerate with $\kappa$. 
This degeneracy is broken by the addition of sources with a redshift-independent distance (anchors); the resulting per-source $H_0$ sensitivity scales as $|1-\beta|$, and vanishes at the calorimetric point.

\begin{figure*}[t!]
\includegraphics[width=0.85\textwidth]{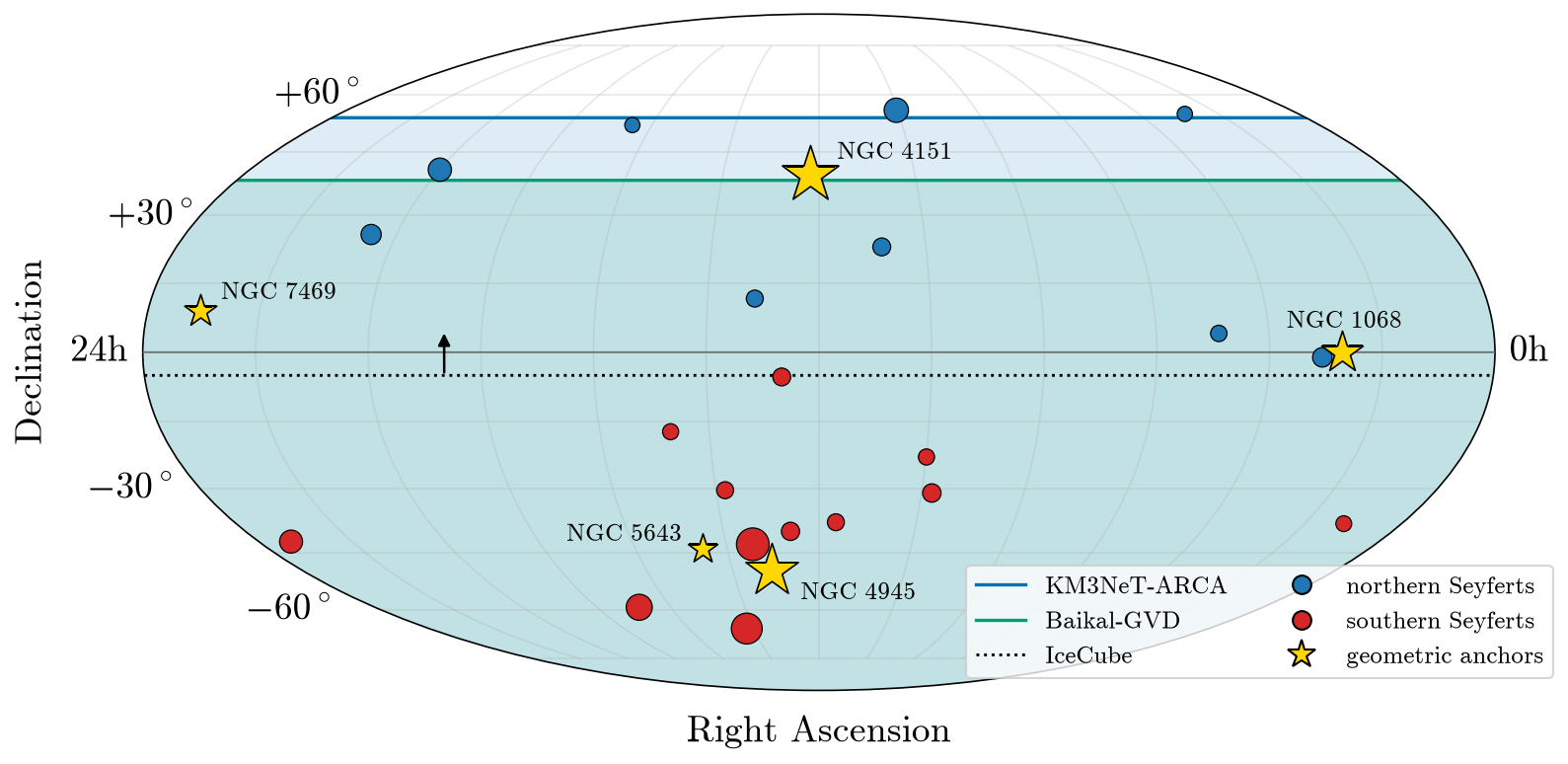}
\caption{
Equatorial (J2000) map of the 26 Seyferts, with marker area proportional to the per-source significance in the 2030 configuration and the five anchors starred. 
Shaded bands show the part of the sky that KM3NeT-ARCA and Baikal-GVD observe as up-going for at least part of the sidereal day; the dotted line and arrow mark IceCube's northern-sky selection.
The southern Seyferts, down-going for IceCube, lie in the up-going sky of both northern detectors.
}
\label{fig:skymap}
\end{figure*}

\begin{figure}[t!]
\includegraphics[width=\columnwidth]{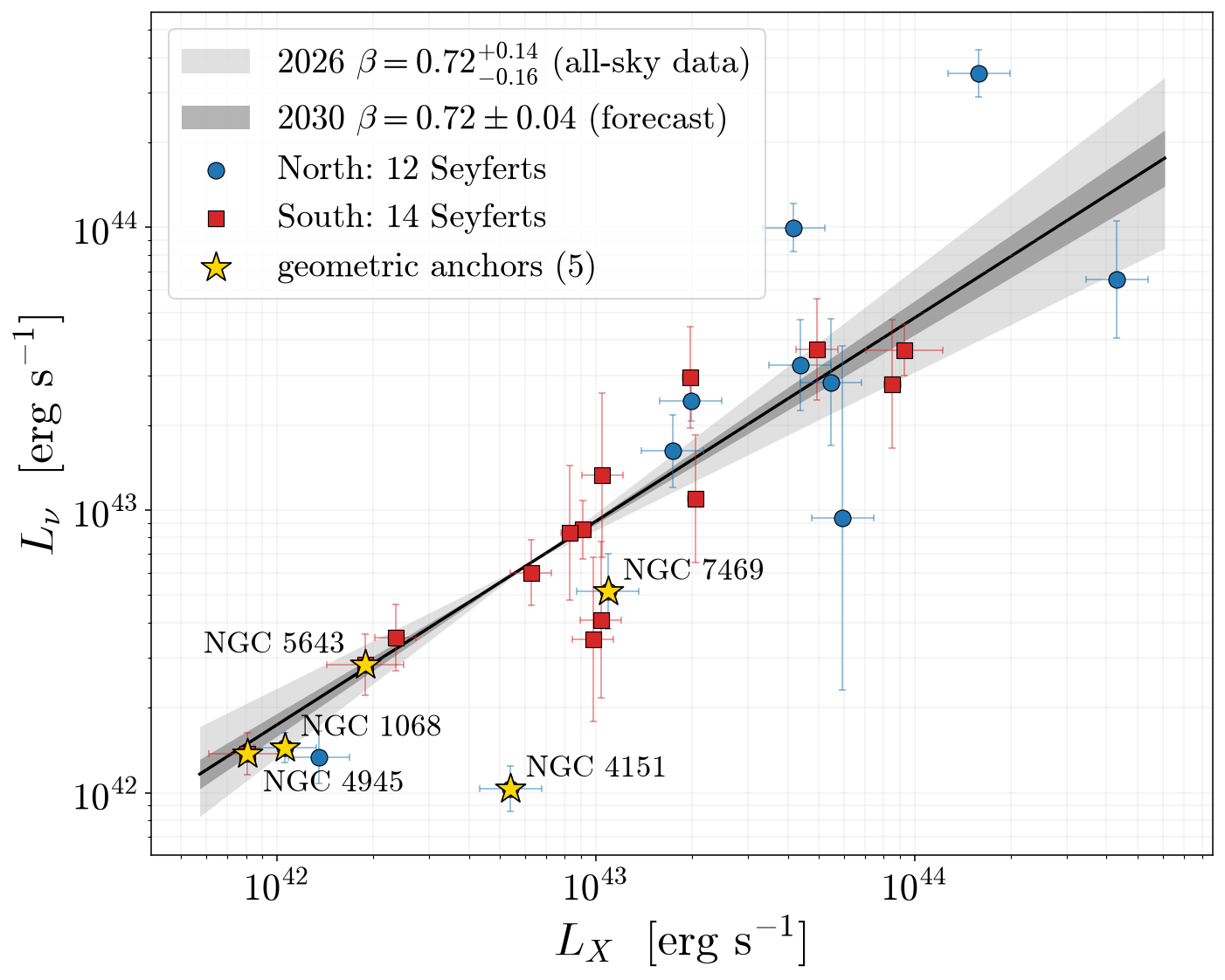}
\caption{
The $L_\nu$--$L_X$ relation in the 2030 projection. 
Northern (blue) and southern (red) Seyferts trace a common power law over more than two decades in luminosity; the black line is the best fit. 
The light band is the present constraint on the slope, $\beta=\betaNow$; the dark band is the 2030 forecast, $\beta=\betaFut$. Error bars indicate the \emph{projected} uncertainties. 
Gold stars mark the five redshift-independent distance anchors used in the 2030 forecast.
}
\label{fig:lxlnu}
\end{figure}

This work analyzes an all-sky sample of 26 Seyfert galaxies, composed of 12 northern sources from the original neutrino--$H_0$ analysis and 14 southern sources selected from IceCube's muon-neutrino southern-sky search~\cite{Abbasi2026ESTES}, see Fig. \ref{fig:skymap} for skymap. 
We then constrain $(H_0,\beta)$ using IceCube's 14-year public data release~\cite{IceCubeDR2}.
Our inference relies on per-source two-dimensional likelihood surfaces $\mathcal{L}(n_s,\gamma)$ which we use to profile over the spectral index $\gamma$. 
This is especially relevant for the low-significance sources, whose spectral index is weakly determined and whose profiles are correspondingly broad. 
To set the absolute scale of the 2026 all-sky fit we select our anchors following the procedure outlined in Ref.~\cite{Herrera:2026now}. 
For the northern sky, we select NGC~1068, whose Cepheid distance is $10.72\pm0.52$~Mpc~\cite{Markham2026NGC1068} and is the most significant northern-sky neutrino source. 
In the southern sky, we select NGC~4945, the most significant southern source and one of the nearest active galaxies, with a tip-of-the-red-giant-branch distance of $3.6\pm0.2$~Mpc~\cite{Mouhcine2005,Tully2015CenA,MouldSakai2008}. 
With these two anchors, the present-day all-sky fit to all 26 sources yields $H_0 = \HzToday$~km~s$^{-1}$~Mpc$^{-1}$ and $\beta = \betaNow$ (Appendix~\ref{app:scans}).
\section{The 2030 forecast for $H_0$ and $\beta$}
\label{sec:h0}
 
Our 2030 prediction sits on several assumptions: a fully complete KM3NeT-ARCA and Baikal-GVD, improved X-ray calibration of our anchors, spectral-index precision on the anchors matching what NGC~1068 already achieves, and a factor-two gain in per-source signal efficiency, representing the historical trend of reconstruction and event-selection improvements to IceCube's point-source sensitivity.
We also assume that the uncertainties on the peculiar velocities of the sources are reduced by a factor of 2.5.

Additionally, we promote three further sources to anchors for a total of five. 
These three additions are NGC~5643 (tip-of-the-red-giant-branch, $13.4\pm0.5$~Mpc~\cite{Hoyt2026TRGB}), NGC~4151 (Cepheid, $15.8\pm0.4$~Mpc~\cite{Yuan2020NGC4151}), and NGC~7469 (Type~Ia supernova SN~2008ec, $58.2\pm1.9$~Mpc~\cite{Koshida2017}).
We promote them in the forecast under the Asimov assumption that they lie on the $L_\nu \propto L_X^\beta$ relation, such that the added exposure
makes them informative anchors. 
We use only redshift-independent and avoid flow-field catalog values, so that the anchor set does not import an assumed value of $H_0$. 
Centaurus~A, the nearest AGN in the sample, carries a precise tip-of-the-red-giant-branch distance but a prominent radio jet, so its neutrino emission need not trace the disk--corona relation, and we exclude it as an anchor. 

To forecast the performance of a global neutrino observatory network, we inject Asimov data on today's best-fit relation ($H_0=66$~km~s$^{-1}$~Mpc$^{-1}$, $\beta=0.72$) and sum each detector's day-averaged up-going response, weighted by their respective exposures. 
With the five anchors and the improvements outlined above, this 2030 fit yields $H_0 = \HzOpt$~km~s$^{-1}$~Mpc$^{-1}$ and $\beta = \betaFut$.

The slope $\beta$ is read from the $L_\nu$--$L_X$ relation of Figure~\ref{fig:lxlnu}, populated across both hemispheres, which the 26 sources constrain to $\beta=\betaNow$ with current data. 
The 2030 values above are a projected precision, assuming the southern Seyferts, which IceCube resolves mostly as down-going upper limits, follow the same relation once KM3NeT-ARCA and Baikal-GVD observe them as up-going tracks. The posteriors of $H_0$ and $\beta$ for both the present-day fit and the forecast are shown in Fig.~\ref{fig:h0}, with the parameters, priors, and sampler detailed in Appendix~\ref{app:scans}.

\begin{figure*}[t]
\includegraphics[width=\textwidth]{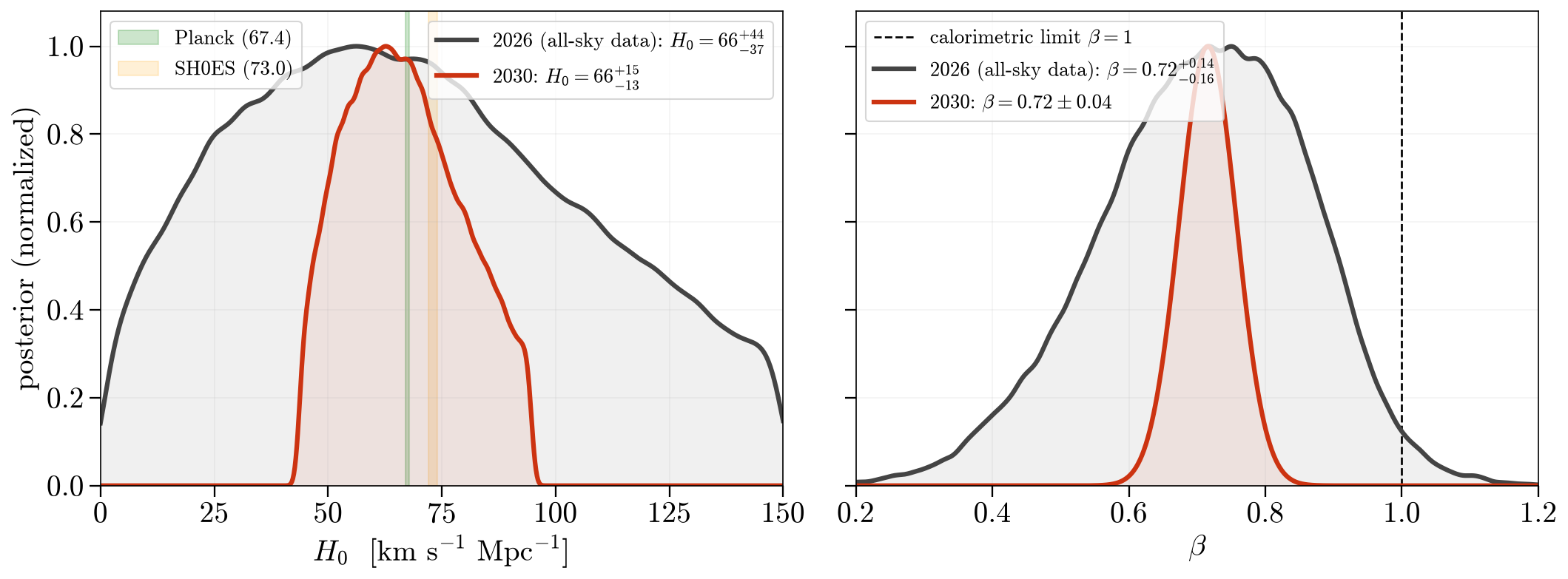}
\caption{
Forecast posteriors, each curve normalized to its peak. (Left) $H_0$: the 2026 all-sky real-data fit (gray, $\HzToday$) and the 2030 configuration (red, $\HzOpt$); Planck and SH0ES bands are shown for reference. 
(Right) The disk--corona slope: the 2026 fit (gray, $\beta=\betaNow$) and the 2030 forecast (red, $\beta=\betaFut$); the dashed line marks the calorimetric limit $\beta=1$. 
}
\label{fig:h0}
\end{figure*}

\section{Adding KM3NeT and Baikal-GVD}
\label{sec:network}

The addition of KM3NeT and Baikal-GVD improves the $H_0$ measurement by significantly improving the sensitivity to southern sky sources. 
For IceCube the southern sky lies above the horizon, where it is buried under the atmospheric-muon background. 
From the northern-hemisphere sites of KM3NeT-ARCA (Mediterranean, $+36^\circ$) and Baikal-GVD (Lake Baikal, $+52^\circ$) the same sky lies below the horizon, and its sources arrive as up-going neutrinos through the Earth, the direction in which atmospheric muon backgrounds are entirely suppressed. 
We thus forecast each detector by mapping IceCube's measured effective area through the detector's real horizon, as in the PLE$\nu$M forecasts of Schumacher et al.~\cite{Schumacher2025PLEnuM}, and scaling by its integrated exposure and a per-unit-exposure track efficiency $\eta$. 
Following Ref.~\cite{Schumacher2025PLEnuM}, ARCA is taken as identical to IceCube per unit volume ($\eta=1$). 
However, we find that the assumption of $\eta=1$ for Baikal-GVD is too strong considering that it is a sparse array of small, optically independent clusters (in contrast to the regular grids of IceCube and KM3NeT-ARCA). 
Thus, we assume a more conservative value, $\eta=0.2$, derived in Appendix~\ref{app:det} from the collaboration's published effective volumes \cite{BaikalTracks}.

With the projected 2030 exposures (IceCube $20$, ARCA $2.3$, Baikal $7.4$ km$^3\cdot$yr; Appendix~\ref{app:det}) the bright southern sources \emph{switch on}.
The up-going acceptance of KM3NeT and Baikal-GVD exceeds IceCube's down-going acceptance in the South by two to three orders of magnitude (the count-error model is given in Appendix~\ref{app:det}). 
As a result, NGC~4945, the strongest southern source today, reaches a per-source signal-to-noise of $\sim\!20$ by 2030, and Centaurus~A, Circinus, and ESO~138-1 reach $7$--$12$ (up from $\lesssim\!0.1$ today), while NGC~5643 and NGC~7582 reach $\sim\!4$--$5$.
By 2030, we assume these sources will be detected and populate the $L_\nu$--$L_X$ relation (Fig.~\ref{fig:lxlnu}), and that NGC~5643 will become a usable calibrator.
The remaining eight southern Seyferts stay at or below $\sim\!2\sigma$ even in 2030. 
The network's leading effect is thus to activate the southern sky and tighten $\beta$; its effect on $H_0$ is indirect, mediated by the anchors it brings online. 
A fuller account of the scaling factors, their uncertainties, and the exposure projections is given in Appendix~\ref{app:det}.

\section{Improving the measurement: X-ray modeling and telescope time}
\label{sec:xray}
Because the anchors break the $\kappa$--$H_0$ degeneracy, an error on their X-ray flux passes directly into $H_0$. 
The sources that matter most for reducing the $H_0$ uncertainty are therefore the three heavily Compton-thick anchors, NGC~1068, NGC~4945, and NGC~5643, whose fluxes are the least certain. 
Most of their hard X-rays are absorbed before reaching us, so their intrinsic fluxes are inferred from absorption models that raise the observed $14$--$195$~keV flux by factors of $6$, $6$, and $17$, respectively~\cite{Ricci2017BASS}.
The larger the correction, the more the result depends on the model. 
Compton-thin sources such as NGC~4151, by contrast, need almost no correction~\cite{Ricci2017BASS}, so their fluxes are limited by calibration rather than modeling~\cite{Madsen2017IACHEC}. 
For the Compton-thick anchors the limit is the modeling itself, since statistical errors are already small for every source except NGC~5643.
About $300$~ks of NuSTAR with simultaneous XMM coverage would constrain the column density and the reflection fraction jointly, the two quantities the de-absorption depends on. 
The deepest pointing would go to NGC~5643, which has the smallest hard-X-ray archive of the anchors and the largest correction. 
The present-day analysis assumes a blanket $\sigma_{F_X}=0.25$ for every source. 
For 2030 we adopt a target of $\sim0.15$ for the Compton-thick anchors, and $\sim0.05$ for Compton-thin sources, in line with what their existing data already support. 
Reaching these targets delivers most of the X-ray gain in $H_0$, while better flux modeling for the non-anchor sources mostly helps tighten $\beta$. 
Appendix~\ref{app:xray} gives the per-class flux budget and the observing plan.

\section{Toward dark energy: high-redshift populations}
\label{sec:de}

High-energy neutrinos can also provide a novel measurement of matter density and dark-energy parameters. 
If a faraway neutrino source population obeys the same $L_\nu$--$L_X$ relation as the Seyferts,
its neutrino flux can be used to extend the corresponding distance ladder to higher redshifts. 
This enlarges the redshift baseline over which the luminosity-distance relation is measured, providing increased sensitivity to the expansion history.
The ``little red dots''~\cite{Akins2025LRD} observed by JWST are abundant, compact active nuclei at $z\sim4$--$7$,  making them a potentially valuable high-redshift extension of the neutrino distance ladder.
As proposed in recent work~\cite{KuzeMurase2026}, these little red dots may supply up to 30\% of the diffuse high-energy neutrino flux.

Extending the Seyfert relation to these sources is less direct. 
Stacked X-ray observations find the little red dots to be faint in X-rays, well below what their broad emission lines would predict for a normal active nucleus~\cite{Yue2024LRDXray}.
As proposed in Ref.~\cite{KuzeMurase2026}, a dense gas envelope may absorb the X-rays while the neutrinos escape, so they can remain bright in neutrinos despite their X-ray faintness. 
A high-redshift neutrino measurement would test this assumption directly. 

Alternatively, it may be possible to anchor the neutrino luminosity of such a source class to a rest-frame optical, UV, or infrared luminosity inferred from JWST observations, through a relation of the form $L_\nu = \kappa'\,L_{\rm obs}^{\beta'}$ that could standardize these sources even when their X-rays are obscured. 
We leave the calibration of such a relation to future work.

\begin{figure}[t]
\includegraphics[width=\columnwidth]{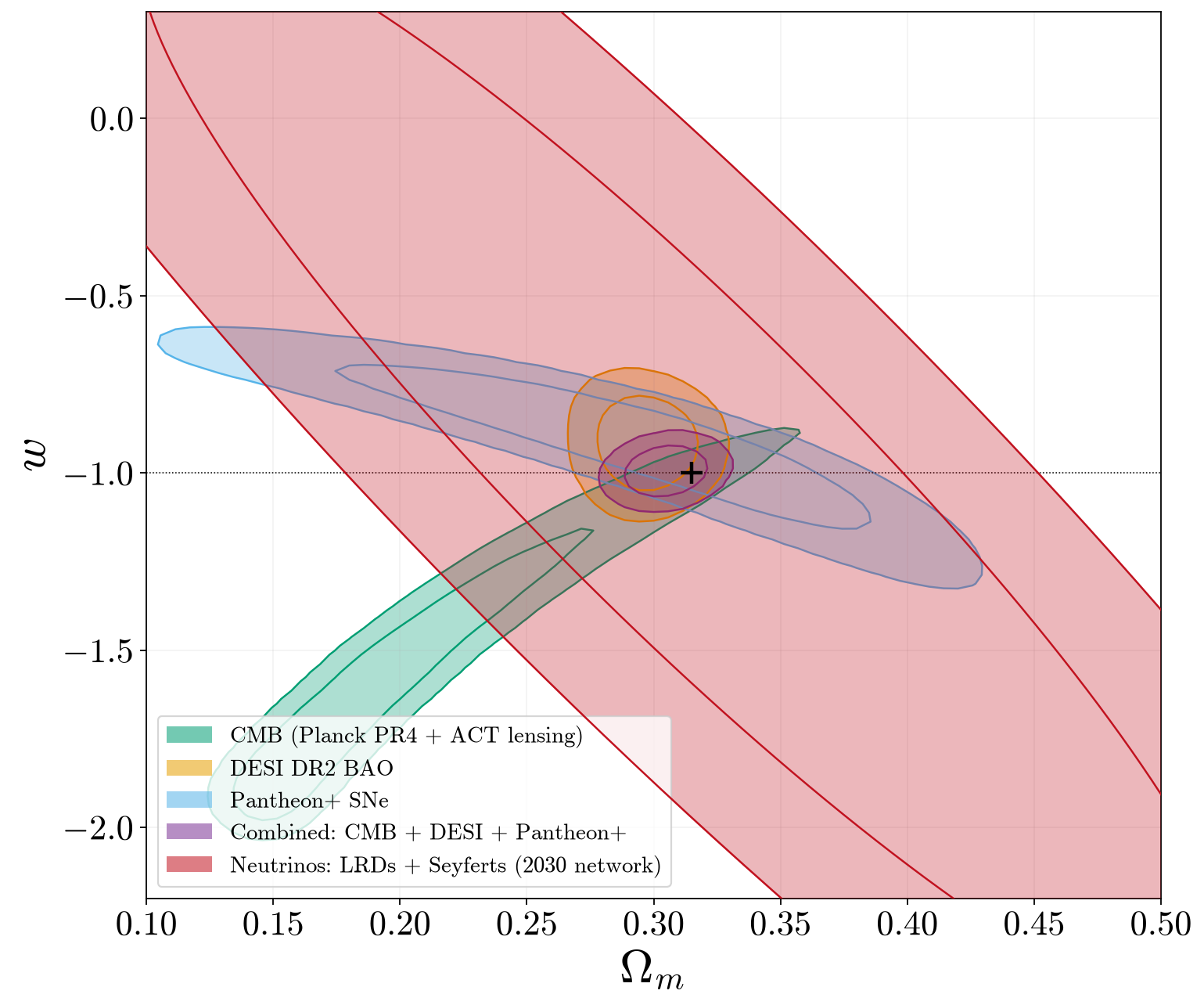}
\caption{
Forecast of $\Omega_m$ and $w$ from a high-redshift neutrino population. 
The neutrino result shows the 68\% and 95\% Fisher band for a population of little red dots spread over $z=4$--$7$, taken to supply $30\%$ of the diffuse neutrino flux~\cite{KuzeMurase2026} and standardized by the Seyferts, whose low-redshift anchors fix $\beta$, $\kappa$, and $H_0$.
The external data contours come from the public chains of DESI DR2~\cite{DESI:2025zgx} and Pantheon+~\cite{Brout:2022vxf}.
}
\label{fig:omegaw}
\end{figure}

Figure~\ref{fig:omegaw} shows a Fisher forecast for the little red dot population.
Following a procedure analogous to that in the Supplemental Material of Ref.~\cite{Herrera:2026now}, we compute the Fisher information on $\Omega_m$ and $w$ using a likelihood that compares the aggregate diffuse neutrino flux contribution of little red dots to the flux predicted from their redshift distribution over $z=4$--$7$~\cite{KuzeMurase2026}.
The low-redshift Seyferts fix the normalization $\kappa$, the slope $\beta$, and $H_0$, so the aggregate high-$z$ flux probes the luminosity distance $d_L(z\!\sim\!5)$ directly.
Folding the northern-tracks diffuse flux~\cite{IceCubeNuMuDiffuse} through the DR2 effective areas and scaling to the 2030 network yields a few thousand events.
Assuming an order-of-magnitude improvement in the flux-normalization and spectral-index uncertainties, the total error on the
population flux is $\approx3\%$. 

This precision produces the region shown in Fig.~\ref{fig:omegaw}.
Marginalizing over the region gives $\sigma(\Omega_m)\approx0.14$ and $\sigma(w)\approx1.0$, so the matter density is pulled to roughly half its prior width while the equation of state stays close to its prior.

The different orientations of the contours in Fig.~\ref{fig:omegaw} reflect the redshift ranges and parameter combinations probed by each observable. In a spatially flat cosmology with constant \(w\), neglecting radiation, the luminosity distance is
\begin{equation}
\begin{split}
d_L(z)&=c\,(1+z)\int_0^z\frac{dz'}{H(z')},\\
H^2(z)&=H_0^2\left[\Omega_m(1+z)^3+(1-\Omega_m)(1+z)^{3(1+w)}\right].
\end{split}
\end{equation}
With the low-redshift calibration and \(H_0\) held fixed, Eq. \ref{eq:model} implies that the constant-flux direction for a source at redshift \(z\) satisfies
\begin{equation}
\left.\frac{dw}{d\Omega_m}\right|_{\Phi_\nu}
=-\frac{\partial\ln d_L/\partial\Omega_m}
{\partial\ln d_L/\partial w},
\end{equation}
with the population constraint involving flux-weighted distance derivatives. Increasing \(\Omega_m\) reduces the distance, while making \(w\) more negative increases it, producing a negative correlation. At the LRD redshifts, the distance integral includes a substantial matter-dominated interval, enhancing its sensitivity to \(\Omega_m\) relative to \(w\), whose influence arises mainly at lower redshifts. This produces a steeper negative tilt than the supernova constraint, which probes the relative distance–redshift relation predominantly at lower redshifts. For illustration, at \(\Omega_m=0.3\) and \(w=-1\), the local constant-distance slope is \(dw/d\Omega_m\simeq-2.2\) at \(z=0.5\), compared with \(-7.2\) at \(z=5\). The CMB instead constrains the acoustic angular scale, \(\theta_*=r_s(z_*)/D_M(z_*)\), together with the physical matter density, \(\omega_m=\Omega_m h^2\), yielding a different geometric degeneracy in which more negative \(w\) is accommodated by larger \(H_0\) and smaller \(\Omega_m\), producing a positive correlation ~\cite{Planck2018}.



\section{Conclusions}
\label{sec:conclusions}

In this work, we find that a global network composed of neutrino telescopes currently either operating or in construction will, by 2030, turn the neutrino distance ladder into an improved probe of the Hubble constant. 
In such a plausible future configuration, $H_0$ sharpens from today's $\HzToday$ to $\HzOpt$~km~s$^{-1}$~Mpc$^{-1}$, a reduction of nearly 65\% in uncertainty.
The disk--corona slope is projected to be pinned to $\beta=\betaFut$, excluding the calorimetric limit at $\sim\!5\sigma$ or greater.
On the neutrino end, these gains are driven by increased northern-sky exposure, activating the southern sky with KM3NeT and Baikal-GVD, and gain in per-source sensitivity from improved reconstruction.
Additionally, the Southern hemisphere could reveal new Seyfert neutrino sources, which would further improve a future $H_0$ measurement.
On the X-ray end, improvements are driven by better flux measurements of a small number of anchor galaxies and more accurate modeling of X-ray absorption at the source.
Looking ahead, if the ladder reaches a high-redshift population, such as JWST's ``little red dots,'' neutrinos could also help constrain the cosmological parameters $\Omega_m$ and $w$. 
Realizing all of these projections will require close coordination between the neutrino and X-ray communities, an effort that would advance both cosmology and our understanding of active-galaxy physics.

\begin{acknowledgments}
We are grateful to Paloma Mustrangi-Carloni and Kiara Carloni for their helpful comments on this manuscript.
This work uses public data from the IceCube Neutrino Observatory 
(14-year point-source release). P.Z. is funded by the David \& Lucile Packard 
Foundation.
The work of G.H. is supported by the Neutrino Theory Network 
Fellowship with contract number 726844. 
 N.K. is supported by the National Science Foundation (NSF, CAREER Grant \#2239795) 
 and the David \& Lucile Packard Foundation. C.A.A. are supported by the Faculty of Arts and 
Sciences of Harvard University, the National Science Foundation (NSF, CAREER Grant \#2239795),
 the John Templeton Foundation (Grant \#63651), 
 the Research Corporation for Science Advancement, 
 and the David \& Lucile Packard Foundation. 
\end{acknowledgments}

\newpage
\section*{AI Policy}
An AI system assisted with the preparation of this manuscript, including drafting portions of the text, developing the analysis, and generating plotting scripts. The scripts drew primarily on code previously developed by the authors and presented in Ref.~\cite{Herrera:2026now}. The authors developed the central ideas, reviewed and verified the code and derivations, and assume full responsibility for the content of this article.

\bibliographystyle{apsrev4-2}
\bibliography{refs}



\clearpage
\newpage

\appendix

\ifx \standalonesupplemental\undefined
\setcounter{figure}{0}
\setcounter{table}{0}
\setcounter{equation}{0}
\fi

\renewcommand{\figurename}{SUPPL. FIG.}
\renewcommand{\tablename}{SUPPL. TABLE}

\renewcommand{\theequation}{A\arabic{equation}}

\section{Likelihood, sampling, and per-source scans}
\label{app:scans}
\begin{figure*}[p]
\includegraphics[width=\textwidth]{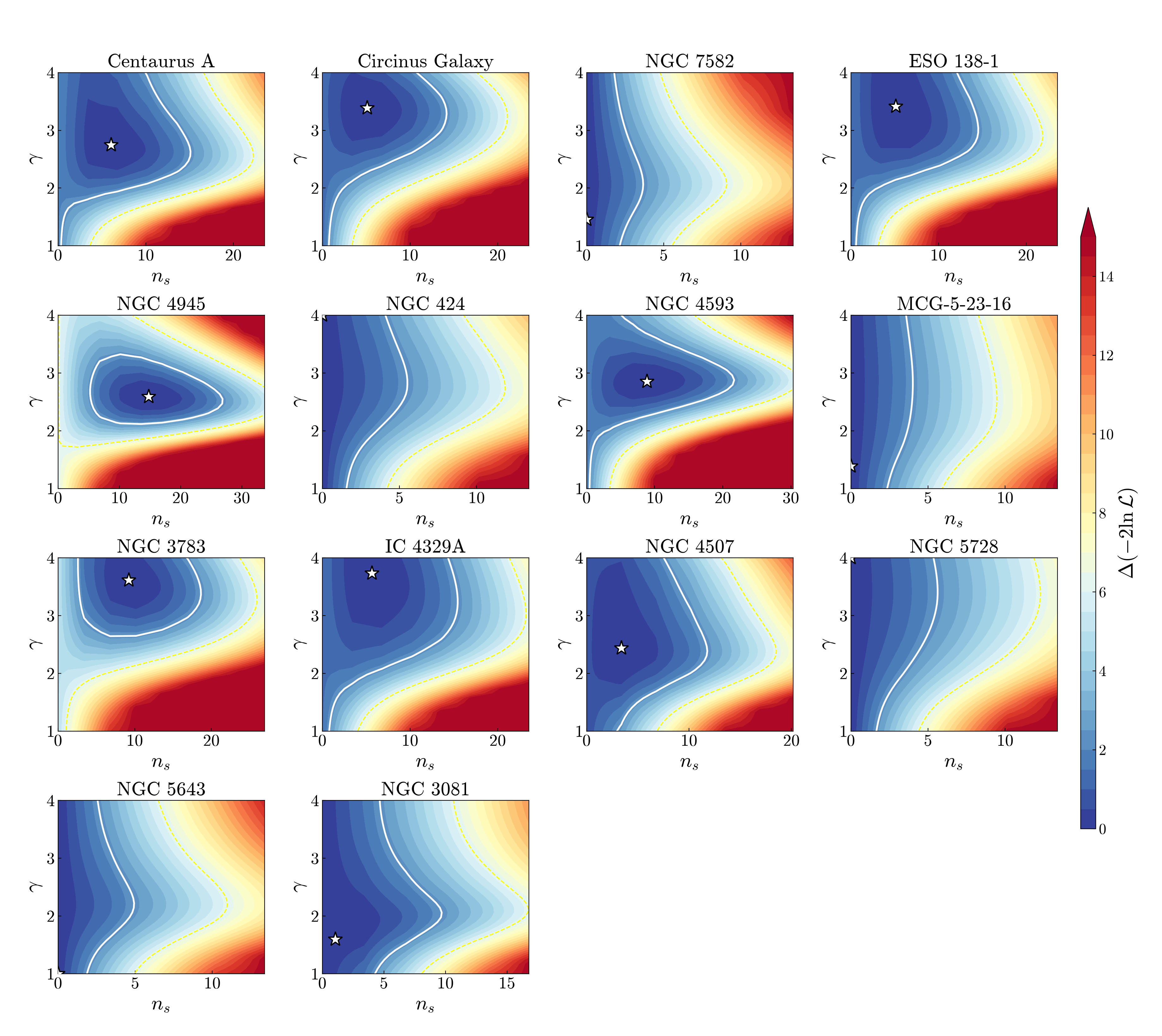}
\caption{Present-day $(n_s,\gamma)$ profile-likelihood scans for the southern Seyferts, one panel each, from IceCube's 14-year public release~\cite{IceCubeDR2}. The star marks the best fit and the white line indicates the 1$\sigma$ confidence level. The northern-sky scans are given in Ref.~\cite{Herrera:2026now}.
}
\label{fig:scansouth}
\end{figure*}

The present-day fit samples the two physical parameters, $H_0$ and $\beta$,
together with the two anchor distances, each with a Gaussian prior from its
distance measurement (Sec.~\ref{sec:method}), under flat priors
$H_0\in(0,150)$~km~s$^{-1}$~Mpc$^{-1}$ and $\beta\in(0.05,3)$. 
At each likelihood evaluation, we profile over the normalization $\kappa$ by first maximizing on a coarse grid in $\ln\kappa$ and then refining the maximum locally. The
per-source spectral index is profiled on the SkyLLH surface (see Figure~\ref{fig:scansouth}): the model count
$\mu_i(\gamma)=\kappa\,C_i(\gamma)\,F_{X,i}^{\beta}\,d_i^{2(\beta-1)}$ is
evaluated on the $\mathcal{L}(n_s,\gamma)$ scan and the minimum of
$\Delta(-2\ln\mathcal{L})$ over $\gamma$ is taken. Following Ref. \cite{Herrera:2026now}, we use a blanket X-ray flux uncertainty of $\sigma_{F_X}=0.25$
and peculiar-velocity uncertainty of $250$~km~s$^{-1}$. 
These terms 
are marginalized by averaging the likelihood over a log-normal spread in
$\mu_i$, evaluated with a five-point Gauss--Hermite quadrature. The posterior
is sampled with a Metropolis--Hastings chain of eight walkers and
$1.6\times10^5$ steps each, discarding the first 30\%; the Gelman--Rubin
statistic satisfies $\hat R<1.01$ for every parameter. 

The forecast keeps a similar structure but with five anchors instead of two, and replaces the scan surfaces with a Gaussian likelihood. In both hemispheres Asimov counts are injected at today's best fit, with per-source variance $\sigma_{n_s}^2+(\sigma_{F_X}\mu)^2+(f_\gamma\mu)^2+(f_v\mu)^2$. 
The count error $\sigma_{n_s}$ combines in quadrature the source's measured 14-year count error grown as the square root of IceCube's exposure, the atmospheric background the up-going detectors add, taken as a constant per unit of up-going acceptance calibrated on the detected northern Seyferts, and the Poisson fluctuation of the injected signal.
The X-ray systematic $\sigma_{F_X}$ is assigned by obscuration class, $0.05$, $0.15$, and $0.30$ from Compton-thin to heavily Compton-thick, with the three Compton-thick anchors improved to $0.15$ by a projected deep-torus campaign. 
The spectral-index width in $f_\gamma=|\partial\ln C/\partial\gamma|\,\sigma_\gamma$ is profiled per source for the detected northern sky and set to a population prior, $\gamma=3.0\pm0.67$ from the measured spread of the northern indices, for the southern nulls, with the anchor galaxies tightened to
$\sigma_\gamma\le0.2$, a precision NGC~1068 already achieves.
The peculiar-velocity term $f_v=2(1-\beta)\,\sigma_v/(cz)$ uses $\sigma_v=100$~km~s$^{-1}$, improved from $250$~km~s$^{-1}$ at present, and $\kappa$ is profiled analytically. 
The sampling is then carried out as in the present-day fit.

\section{KM3NeT and Baikal-GVD detector scaling}
\label{app:det}

We model the KM3NeT-ARCA and Baikal-GVD acceptances following the PLE$\nu$M global-network forecasts~\cite{Schumacher2025PLEnuM}. 
Each detector is assigned IceCube's effective area for up-going events, where the Earth filters out the atmospheric-muon background. 
Over a sidereal day, a source at declination $\delta$ moves through each detector's local sky and its zenith angle $\theta_z$ changes. 
While the source is below the horizon, we evaluate IceCube's effective area at the equivalent declination $\sin\delta_{\rm IC} = -\cos\theta_z$. 
Above the horizon we set the acceptance to zero. 
Sources there lie in the northern sky, which IceCube already covers with its greater up-going acceptance. 
Integrating over the source spectrum and averaging over the day gives the per-source acceptance $C_d(\delta,\gamma)$. 
For IceCube itself we use the measured effective area at the true declination, including its weak down-going acceptance in the southern sky. 
Southern sources are down-going at the South Pole but spend most of the day below the horizon at the Mediterranean and Lake Baikal sites, which is where the network gains most of its sensitivity for that hemisphere. 
The network acceptance is the sum of the three $C_d$, each weighted by its exposure and by $\eta$.


The exposures in Table~\ref{tab:det} integrate each detector's instrumented
volume over time, in the volume convention each collaboration uses. 
For IceCube we count full-detector-equivalent years: the 14-season release used
for the calibration (2008--2022, 13.6~yr of livetime) corresponds to 12.7 such
years once the 40-, 59- and 79-string seasons are down-weighted by their
fractional size (0.47, 0.69, and 0.92 of the complete 86-string detector), and 20 accrue by
2030, the value also used in Ref.~\cite{Schumacher2025PLEnuM}. 
For KM3NeT-ARCA we start from the exposure accumulated by mid-2026, 0.36~km$^3\cdot$yr, and grow the instrumented mass linearly to the
complete two-block detector (1~km$^3$) by 2030, giving 2.3~km$^3\cdot$yr. 
For Baikal-GVD we start from 4.0~km$^3\cdot$yr accumulated by mid-2026 with
0.85~km$^3$ instrumented, in the per-cluster convention discussed below, and
grow to 1~km$^3$ by the end of 2027, constant thereafter, giving
7.4~km$^3\cdot$yr. The completion dates are the collaborations' stated targets.

For KM3NeT-ARCA we adopt $\eta=1$, following Ref.~\cite{Schumacher2025PLEnuM},
assuming its instrumented volume behaves roughly like IceCube's per unit volume.
Baikal-GVD, however, is not treated the same, because of the sparsity of its design: 
it is built from optically independent clusters that, although together they span
roughly a cubic kilometer, each act as a standalone detector separated by large
gaps. Tracks are reconstructed within a single small cluster, so
its effective volume for tracks is far smaller than the nominal cubic kilometer.
We therefore adopt $\eta=0.2$, obtained by converting Baikal's published per-cluster effective volumes~\cite{BaikalTracks} to effective areas for the full 20-cluster array and taking their ratio to IceCube's published effective area~\cite{IceCubeDR2} decade by decade in energy. 
The ratio rises from $\sim0.1$ at TeV energies to $\sim0.3$ near a PeV; we chose $\eta=0.2$ as the average of these two values.  

\begin{table}[t]
\caption{2030 integrated exposures (km$^3\cdot$yr), and per-unit-exposure
track efficiency $\eta$ relative to IceCube's northern sky.}
\label{tab:det}
\begin{ruledtabular}
\begin{tabular}{lccc}
Detector & Latitude & Exposure (2030) & $\eta$ \\
\colrule
IceCube      & $-90^\circ$ & $20$  & $1.0$ \\
KM3NeT-ARCA  & $+36^\circ$ & $2.3$   & $1.0$ \\
Baikal-GVD   & $+52^\circ$ & $7.4$  & $0.2$ \\
\end{tabular}
\end{ruledtabular}
\end{table}


\newpage
\section{X-ray flux budget and observing program}
\label{app:xray}

The blanket $\sigma_{F_X}=0.25$ splits, for each source, into four terms. 
They are photon statistics, the de-absorption correction from the absorption or torus model, the mismatch between snapshot X-ray fluxes and the 14-year neutrino average, and the conversion between energy bands. 
Scaling the 105-month Swift/BAT $14$--$195$~keV errors~\cite{Oh2018BAT} to the $20$--$50$~keV band gives 90\% intervals of $0.4\%$ to $33\%$, above $15\%$ for only five sources. 
Averaging the full 2004--2026 archive should bring every source except NGC~5643 below $15\%$. 
We estimate mean floors for the remaining terms of $\sim0.05$ for Compton-thin sources, $\sim0.15$ for mildly Compton-thick sources with $N_{\rm H}$ known to $\sim\!0.1$~dex, and $\sim0.3$ for extreme Compton-thick sources with archival data alone, where the choice of de-absorption model dominates. 
The extreme Compton-thick floor shows that the blanket $0.25$ is already optimistic for the three Compton-thick anchors, which the campaign below aims to bring to $0.15$.

Because an anchor's X-ray flux error passes directly into $H_0$, the observing program can be small and focused. 
New telescope time pays off only where the intrinsic flux is limited by both the model and the data, like NGC~5643. 
Its spectrum is reflection-dominated, and fits to its first $42$~ks of NuSTAR data give intrinsic luminosities of $0.8$ to $1.7\times10^{42}$~erg~s$^{-1}$, a $\sim\!30\%$ spread around the mean driven mainly by slab versus torus reflection geometry~\cite{Annuar2015NGC5643}. 
This spread motivates the extreme Compton-thick floor above. 
About $100$~ks of NuSTAR with simultaneous XMM coverage would cut its statistical error by about a third and constrain the column density and reflection fraction jointly. 
The other anchors need reanalysis rather than new data. 
NGC~1068 already has deep data, and its intrinsic flux is limited by the geometry of its multi-component reflector~\cite{Bauer2015NGC1068}. We assume that a uniform torus fit to this archive, anchored to the BAT average, brings it to the $0.15$ target. NGC~4945 is transmission-dominated above 10~keV~\cite{Puccetti2014NGC4945,Boorman2026NGC4945}, but a line-of-sight Compton depth of $\tau_{\rm T}\approx3$ still leaves its intrinsic flux model-dependent~\cite{Boorman2026NGC4945}, and its hard flux varies by a factor of two within hours. NGC~4151 is Compton-thin and limited only by calibration. 
For these anchors the remaining work is to average archival Swift/BAT fluxes over the neutrino livetime and to fit one physically motivated torus model to every obscured source. 
A further $25$--$40$~ks of NuSTAR on each of the six sample sources with little archival data would make the sample uniform, which would help the $\beta$ measurement.  
In total, about $300$~ks of targeted hard-X-ray time, combined with this reanalysis of existing data, captures most of the X-ray improvement available to $H_0$.

\clearpage

\end{document}